%% file: main.tex
\documentclass{article}

\input{math_commands.tex}

\usepackage{arxiv}

\usepackage[utf8]{inputenc} % allow utf-8 input
\usepackage[T1]{fontenc}    % use 8-bit T1 fonts
\usepackage{hyperref}       % hyperlinks
\usepackage{url}            % simple URL typesetting
\usepackage{booktabs}       % professional-quality tables
\usepackage{amsfonts}       % blackboard math symbols
\usepackage{nicefrac}       % compact symbols for 1/2, etc.
\usepackage{microtype}      % microtypography
\usepackage{lipsum}		% Can be removed after putting your text 
\usepackage{doi}

\usepackage{graphicx}
\usepackage{subcaption}
\usepackage[utf8]{inputenc} % allow utf-8 input
\usepackage[T1]{fontenc}    % use 8-bit T1 fonts
\usepackage{booktabs}       % professional-quality tables
\usepackage{amsfonts}       % blackboard math symbols
\usepackage{nicefrac}       % compact symbols for 1/2, etc.

\usepackage{amsmath}
\usepackage{amssymb}
\usepackage{mathrsfs}
\usepackage{dsfont}
\usepackage{framed}

\usepackage{graphicx}

\usepackage[numbers]{natbib}
\PassOptionsToPackage{compress}{natbib}
\usepackage{hyperref,xcolor}
\hypersetup{
colorlinks,
linkcolor={red!50!black},
citecolor={blue!50!black},
urlcolor={blue!70!black}
}

\usepackage{framed,color}
\definecolor{shadecolor}{rgb}{0.85,0.85,0.85}
\usepackage{ulem}

\title{Principal Trade-off Analysis}

\author{ 
	\href{https://orcid.org/0000-0001-6618-631X}{\includegraphics[scale=0.06]{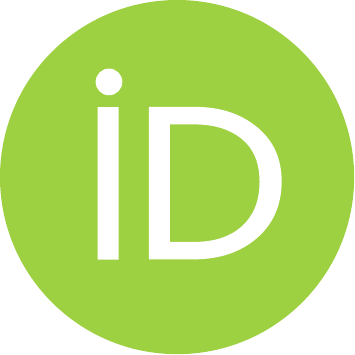}\hspace{1mm}Alexander Strang}\thanks{\href{alexanderstrang.com}{alexanderstrang.com}, Equal Contribution} \\
	Department of Statistics\\
	University of Chicago\\
	Chicago, IL 60637 \\
	\texttt{alexstrang@uchicago.edu} \\
	\And 
	David SeWell 
	\thanks{Equal Contribution} \\
	Lockheed AI Center (LAIC) \\ 
	Lockheed Martin Corp \\
	\texttt{david.r.sewell@lmco.com} \\
	\And
    Alexander Kim 
    \thanks{alexander.kim@kbra.com} \\
    Lockheed AI Center (LAIC) \\ 
	Lockheed Martin Corp \\
    \texttt{alexander.kim@lmco.com} \\
	\And
	Kevin Alcedo \\
    Lockheed AI Center (LAIC) \\ 
	Lockheed Martin Corp \\
    \texttt{kevin.alcedo@lmco.com} \\
    \And 
    David Rosenbluth 
    \thanks{Corresponding author}\\
    Lockheed AI Center (LAIC) \\ 
	Lockheed Martin Corp \\
    \texttt{david.rosenbluth@lmco.com} 
}

\date{}

\begin{document}

\maketitle

\begin{abstract}
      %The focus on equilibrium solutions in games underemphasizes the importance of understanding their overall structure. A different set of tools is needed for learning and representing the general structure of a game. 
      How are the advantage relations between a set of agents playing a game organized and how do they reflect the structure of the game?
      In this paper, we illustrate "Principal Trade-off Analysis" (PTA), a decomposition method that embeds games into a low-dimensional feature space. We argue that the embeddings are more revealing than previously demonstrated by developing an analogy to Principal Component Analysis (PCA).  PTA represents an arbitrary two-player zero-sum game as the weighted sum of pairs of orthogonal 2D feature planes. We show that the feature planes represent unique strategic trade-offs and truncation of the sequence provides insightful model reduction.  We demonstrate the validity of PTA on a quartet of games (Kuhn poker, RPS+2, Blotto, and Pokemon). In Kuhn poker, PTA clearly identifies the trade-off between bluffing and calling. In Blotto, PTA identifies game symmetries, and specifies strategic trade-offs associated with distinct win conditions. These symmetries reveal limitations of PTA unaddressed in previous work. For Pokemon, PTA recovers clusters that naturally correspond to Pokemon types, correctly identifies the designed trade-off between those types, and discovers a rock-paper-scissor (RPS) cycle in the Pokemon generation type - all absent any specific information except game outcomes. 
\end{abstract}


\section{Introduction}

In recent years algorithms have achieved superhuman performance in a number of complex games such as Chess, Go, Shogi, Poker  and Starcraft \cite{silver2018general,heinrich2016deep, moravvcik2017deepstack,Grandmaster}. Despite impressive game play, enhanced understanding of the game is typically only achieved by additional analysis of the algorithm's play post facto \cite{chess_new_light}. Current work emphasizes the ``policy problem", developing strong agents, despite growing demand for a task theory which addresses the ``problem problem", i.e.~what games are worth study and play \cite{omidshafiei2020navigating, clune2019ai}. A task theory requires a language that characterizes and categorizes games, namely, a toolset  of measures and visualization techniques that evaluate and illustrate game structure. Summary visuals and measures are especially important for complex games where direct analysis is intractable. In this vein, tournaments are used to sample the game and to empirically evaluate agents. The empirical analysis of tournaments has a long history, in sports analytics \cite{lewis2004moneyball, bozoki2016application} , ecology and animal behavior \cite{Laird, silk1999male}, and biology \cite{stuart2006multiple, Sinervo}. While the primary interest in these cases is typically in ranking agents/players, tournaments also reveal significant information about the nature of the underlying game \cite{tuyls2018generalised}. This paper describes mathematical techniques for extracting useful information about the underlying game structure directly from tournament data. While these methods 
can be applied to the various contexts in which tournaments are already employed in machine learning (e.g., population based training), they open up a range of new research questions regarding the characterization 
of natural games, synthesis of artificial games (c.f.~\cite{omidshafiei2020navigating}), game approximation via simplified dynamics, and the strategic perturbation of games.

Note that our objectives are as empirical as they are game theoretic.  Empirical game theory, the study of games from actual game play data (e.g. sports analytics), studies games \textit{as they are played} by a particular population, rather than by an idealised player. Thus, empirical game theory has its own, valid, objectives beyond finding equilibria or optimal players. Exclusive focus on optima ignores the global structure of a game as it is experienced by the majority of participants. What decision dilemmas do they face? What game dynamics do they experience? What game space must they navigate in the process of optimization? How should they exploit a chosen opponent, population, or form teams? All of these questions are more easily addressed given a simplified global representation that isolates each important independent aspect of a game. PTA offers such a summary.

PTA represents structural characteristics of a tournament by a low dimensional embeddings that maps competitive relationships to embedded geometry. We review and expand on methods introduced by \cite{balduzzi2018re}, who proposed a series of maps that describe a sample tournament in terms of a sum of simple games, namely, disc games. 

Our contribution follows. First, we compare PCA \cite{pearson1901liii} to disc game embedding, and show that disc game embeddings inherit the key algebraic properties responsible for the success of PCA. Based on this analogy, we propose PTA as a general technique for visualizing data arising from competitive tasks or pairwise choice tasks. Indeed, while we focus on games for their charisma, any data set representing a skew-symmetric comparison of objects is amenable to PTA. Via a series of examples, we show that PTA provides a much richer framework for analyzing trade-offs in games than previously demonstrated. Our examples exhibit a wide variety of strategic trade-offs that can be clearly visualized with PTA. Unlike existing work, we focus on the relation between embedding coordinates, which represent performance relations, and underlying agent attributes, to elucidate the principal trade-offs responsible for cyclic competition. Moreover, we consider the full information content of PTA by analyzing multiple leading disc games and the decay in their importance. Important strategic trade-offs can arise in later disc games, so previous empirical work's focus on the leading disc game is myopic. These examples also raise conceptual limitations not addressed in previous work, thus outline future directions for development.

\section{Related Work}
\label{Related Work}
Our work builds directly on \cite{balduzzi2018re}, which used the embedding approach to introduce a comprehensive agent evaluation scheme.  
Their scheme uses the real Schur form (PTA) in conjunction with the Hodge decomposition to overcome deficiencies in standard ranking models. Our work also compliments efforts to explore cyclic structures in competitive systems \cite{Candogan,HHD}, economics \cite{linares2009inconsistent, may1954intransitivity}, and tangentially as multi-class classification problems \cite{bilmes2001intransitive, huang2006generalized}. Cycles challenge traditional gradient methods and can slow training \cite{omidshafiei2020navigating, balduzzi2018mechanics}. Accordingly, highly cyclic games, such as iterated rock-paper-scissors, are useful as benchmark tasks for mulit-agent reinforcement learning \cite{lanctot2023population}. Moreover, cyclic structures in games are often intricate and difficult to disentangle, particularly among intermediate competitors. Games of skill frequently exhibit this ``spinning top geometry" \cite{czarnecki2020real}. By summarizing cyclic structures, PTA helps identify areas of the strategy space that cause difficulty during training, or should be targeted for diverse team design \cite{balduzzi2019open, garnelo2021pick}. Here, we show that PTA can identify fundamental trade-offs that summarize otherwise opaque cyclic structure. Trade-offs play an important role in decision tasks and evolutionary processes outside of games, so general tools that isolate and reify trade-offs are of generic utility \cite{omidshafiei2020navigating, tuyls2018generalised}. In that sense, our attempt to visualize game structure is in line with generic data visualization efforts, which aim to convert complicated data into elucidating graphics (c.f.~\cite{healy2018data, garnelo2021pick}). 

\section{Background}
\label{Background}

\subsection{Functional Form Games}
A two-player zero-sum functional form game, is defined by an attribute space $\Omega \subseteq \mathbb{R}^{T}$ and an evaluation function $f$ that returns the advantage of one agent over another given their attributes. Agents in the game can be represented by their attribute vectors $x,y \in \Omega$, whose entries could represent agent traits, strategic policies, weights in a neural net governing their actions, or more generally, any attributes that influence competitive behavior.  The function $f$ is of the form $f: \Omega \times \Omega \rightarrow \mathbb{R}$. The value $f(x,y)$, quantifies the advantage of agent $x$ over $y$ with a real number. We assume that advantage is zero-sum. Consequently, $f$ must be skew-symmetric, $f(x,y) = -f(y,x)$ \cite{HHD}. If $f(x,y) > 0$ we say
that $x$ beats $y$ and the outcome is a tie if $f(x,y) = 0$. The larger $|f(x,y)|$, the larger the advantage one competitor possesses over another. We do not specify how advantage is measured, since the appropriate definition may depend on the setting. Possible examples include expected return in a zero-sum game, probability of win minus a half, or log odds of victory. With a set of agents $X$, pairwise comparisons of all agents gives a $N \times N$ evaluation matrix $F$ where $F_{ij} = f(x_{i},x_{j})$. Any such matrix can be separated into transitive and cyclic components, $F_t$ and $F_c$, via the Helmholtz-Hodge decomposition (HHD) \cite{balduzzi2018re,HHD,lim2020hodge}. The HHD writes $F = F_t + F_c$ where $[F_t]_{ij} = r_i - r_j$, and where $r$ are least squares ratings that evaluate the average performance of each agent. These matrices can be analyzed to study the structure of the game among those competitors, i.e.~the resulting tournament. 

\subsection{Disc Games}

The cyclic component of a tournament can be visualized using a combination of simple cyclic games \cite{balduzzi2019open,balduzzi2018re}. The simplest cyclic functional form game is a disc game, which acts as a continuous analog to rock-paper-scissors (RPS) in two-dimensional attribute spaces. The disc game evaluation function is the cross product between competitor's embedded attributes, 
\begin{equation} \label{eqn: disc}
  \text{disc}(x,y) = x \times y = x_1 y_2 - x_2 y_1 =  x^{T} \begin{bmatrix} 0 & 1 \\
    -1 &  0  
\end{bmatrix} y = x^{T} R y
\end{equation}
where $R$ is the 2 $\times$ 2 ninety degree rotation matrix \cite{balduzzi2018re}. The cross product models a basic trade-off between the two attributes.

Any evaluation matrix can be represented with a sum of pointwise embeddings into a sequence of disc games. The necessary construction follows.

\section{Principal Trade-off Analysis (PTA)}
\label{PTA}

PTA decomposes an arbitrary performance matrix $F$ into a sum of simpler performance matrices by embedding each agent into a series of disc games that model important strategic trade-offs. 

Any real, $m \times m$, skew-symmetric matrix $A$ admits a Schur decomposition (real Schur form),  $QUQ^{T}$. Here $Q$ is an orthonormal $m \times \text{rank}(A)$ matrix, $U$ is block diagonal with $\text{rank}(A)/2$, $2 \times 2$ blocks of the form $ U^{(k)} = \omega_{k} R$, where $\omega_k \geq 0$ is a nonnegative real number. Each pair of consecutive columns of $Q$, $[q_{2k-1},q_{2k}]$, correspond to the real and imaginary parts of an eigenvector of $A$ scaled by $\sqrt{2}$. The scalars $\omega$ are the nonnegative imaginary part of the corresponding eigenvalues, listed in decreasing order \cite{youla1961normal, zumino1962normal}. A linear algebra exercise demonstrates that the columns of $Q$ are also proportional to the singular vectors of $A$, and the sequence of scalars $\omega$ match the singular values of $A$. A truncated expansion of $A$ using only the first $r$ blocks is equivalent to truncating the singular value decomposition at $2r$ singular vectors, so, by the Eckart-Young-Mirsky theorem, equals the optimal rank $2r$ approximation to $A$ under the Frobenius norm \cite{eckart1936approximation,mirsky1960symmetric,strang2019linear}.

When $A$ is replaced with the performance matrix $F$, each block in the Schur decomposition acts as a scaled version of a disc game where each competitor is assigned embedding coordinates via $Q$. The performance matrix $F$ is skew-symmetric, so admits a Schur decomposition:
\begin{equation} \label{eqn: Schur}
F = QUQ^{T}.
\end{equation}

As in PCA, we consider a low-rank approximation of $F$ associated with expansion onto the first $r$ disc games, where $r$ is chosen large enough to satisfy a desired reconstruction accuracy. Low-rank approximation allows model reduction and mixed equilibria approximation in quasi-polynomial time \cite{lipton2003playing}. The closest rank $2r$
approximation to $F$ in Frobenius norm is given by replacing $Q$ with $Q^{(1:2r)}$, and $U$ with $U^{(1:2r)}$ in Equation \ref{eqn: Schur}, where $Q^{(1:2r)}$ is the first $2r$ columns of $Q$, and $U^{(1:2r)}$ is the upper $2r$ by $2r$ minor of $U$.  As in other latent variable models , the feasibility of low-rank approximation can be justified when $f$ is sufficiently smooth via the Johnson-Lindenstrauss lemma \cite{udell2019big}. %Optimality is measured using Frobenius norm error. 

The matrix $Q^{(1:2r)}$ provides a set of basis vectors. Projection onto those basis vectors defines a new set of coordinates, thereby embedding the competitors. Specifically, let:
\begin{equation} \label{eqn: project F_c}
\hat{Y} = Q^{(1:2r)^T} F = U^{(1:2r)}{Q^{(1:2r)^T}}.
\end{equation}
 and scale each pair of embedding coordinates by the associated eigenvalue so $\vec{y}_k(i) = [y_{i,2k-1},y_{i,2k}] = \omega_k^{-1/2} [\hat{y}_{i,2k-1},\hat{y}_{i,2k}] = {\omega_k}^{1/2} [q_{i,2k-1},q_{i,2k}]$. Then $\vec{y}_k(i)$ maps from competitor indices, $i$, to points in $\mathbb{R}^2$, and the set $Y = \{\vec{y}_k\}$ is a collection of planar embeddings, where $\vec{y}_k$ is given by projection onto a feature plane spanned by $q_{2k-1}$ and $q_{2k}$.
 
A user interested in the transitive and cyclic components of $F$ separately, could begin by breaking $F$ into $F_t$ and $F_c$ \cite{HHD}. The transitive component can be represented on a line via the ratings, so does not require additional visualization \cite{balduzzi2018re}. The cyclic component $F_c$ is skew-symmetric, so can be represented via PTA. Then performance is represented by a combination of two components. The first compares the overall quality of the agents, as quantified by a set of ratings. The second represents all remaining cyclic relations as a combination of principal trade-offs.  %We illustrate both approaches in our examples.
  
 Since PTA depends only on $F$, the cost of performing PTA is independent of the complexity of the underlying game or agents. Once $F$ is formed, PTA proceeds at the same cost as standard low-rank matrix decompositions like PCA which seek the leading singular vectors of a symmetric matrix ($\mathcal{O}(n^3)$ for $n$ sampled agents \cite{golub2013matrix}). While iterative algorithms may be more efficient when only a few leading columns of $Q$ are required, the computational cost of performing PTA in an empirical setting will almost always be swamped by the cost of gathering the data for forming $F$, which requires training and comparing $\mathcal{O}(n^2)$ pairs of agents. The simulation cost could be reduced if low-rank completion methods were applied to fill in missing data \cite{meka2009guaranteed,gleich2011rank,chen2016modeling}. We leave in-depth sampling considerations and matrix completion methods to future work.

Note that the Schur decomposition is only unique up to rotation within each feature plane, since complex conjugate pairs of eigenvectors of $F$ are only uniquely defined up to their complex phase. Thus, two embeddings are equivalent if they agree up to rotation within each planar embedding.

The evaluation $F_{ij}^{(2k)}$ between agent $i$ and agent $j$ equals a sum over each embedding $\vec{y}_k$, of the cross product $\vec{y}_k(i) \times \vec{y}_k(j)$ (see Appendix \ref{appendix:pta}). That is:
\begin{equation}
F_{ij}^{(2r)} =  \sum_{k=1}^{r} \vec{y}_k(i) \times \vec{y}_k(j) = \sum_{k=1}^{r} \text{disc}(\vec{y}_k(i),\vec{y}_k(j)).
\end{equation}

Thus, restricted to each embedding $F^{(2r)}$ is a disc game and the optimal rank $2r$ approximation of $F^{(2r)}$ is a linear combination of disc games applied to the sequence of planar embeddings $\{\vec{y}_k\}_{k=1}^{r}$. 

This decomposition is useful for two reasons. First, it depends on a spectral decomposition of $F$, so inherits the key algebraic properties that account for the success of PCA. An equivalent construction is introduced in \cite{chen2016modeling} where it is called the ``blade-chest-inner" model. The construction in \cite{chen2016modeling} is not based on a spectral decomposition, so lacks orthogonality or low-rank optimality. In PTA, the embeddings are projections onto orthogonal planes, so each embedding encodes independent information about cyclic competition. Specifically, the embedding coordinates are uncorrelated when an agent is sampled randomly from the population. The planes act like feature vectors as they are typically associated with some strategic trade-off (see Section \ref{Experiments}). Therefore, as PCA identifies principal components, PTA identifies principal trade-offs: orthogonal planes associated with a sequence of fundamental cyclic modes. The two decompositions differ since PCA uses the singular value decomposition, while PTA uses the Schur real form. Nevertheless, the sequence of embeddings form optimal low-rank approximations to $F$, where the importance of each embedding is quantified by the magnitude of the associated eigenvalue. Thus, the sequence of eigenvalues determines the number of disc game embeddings, $r$, required to achieve a sufficiently accurate approximation of $F$. The number of disc games is half the numerical rank of $F$ and is a natural measure of the complexity of cyclic competition. The complexity is distinct from the overall intensity of cyclic competition or the game balance, which depend on $\|F_c\|$ and $\|F\|$ instead of its rank \cite{HHD}. Instead, it counts the number of distinct cyclic modes in the
evaluation matrix. PTA is most effective when $r$ is small. The smaller $r$, the simpler the associated representation. In our experiments, the numerical rank is typically small and grows slowly in the size of the underlying game or strategy space.

%Second, the disc game construction encodes performance relations via geometry. Given coordinates $\vec{y}(i)$ and $\vec{y}(j)$, the advantage of competitor $i$ over competitor $j$ given by $\vec{y}(i) \times \vec{y}(j)$ equals twice the signed area of the triangle with vertices at the origin, $\vec{y}(i)$, and $\vec{y}(j)$. In polar coordinates, each point in a disc game has a radius and angle. The cross product $\text{disc}(\vec{y}(i),\vec{y}(j))$ equals the product of the radii, times the sine of the angle between $\vec{y}(i)$ and $\vec{y}(j)$. So, the farther a competitor is embedded from the origin, the more intensely they are involved in the associated cyclic mode. For a fixed radius, one competitor gains the most advantage when it is embedded ninety degrees clockwise from its opponent, and possesses an advantage as long as it is embedded clockwise of its opponent. Thus, advantage flows clockwise about the origin. We visualize this flow with a circulating vector field. These geometric properties allow the sequence of disc games to encode a variety of cyclic structures in interpretable visuals. Our subsequent analysis relies heavily on these properties. 

Second, disc games encode performance in geometry. Given coordinates $\vec{y}_k(i)$ and $\vec{y}_k(j)$, the advantage competitor $i$ possesses over $j$ is given by $\vec{y}_k(i) \times \vec{y}_k(j)$.
In polar coordinates, each agent is assigned a radius and angle per disc game, $(r_k(i),\theta_k(i))$. Then, $\text{disc}(\vec{y}(i),\vec{y}(j)) = r_k(i) r_k(j) \sin(\theta_k(j) - \theta_k(i))$. The larger $r_k(i)$, the more the $k^{th}$ cyclic mode influences the performance of agent $i$. For a fixed radius, one competitor gains the most advantage when it is embedded $90^{\circ}$ clockwise from its opponent and possesses an advantage as long as it is embedded clockwise of its opponent. Thus, advantage flows clockwise about the origin. We visualize this flow with a circulating vector field, $\vec{v}(\vec{y}) = [y_2,-y_1]$. This intuitive geometry allows disc games to encode a variety of cyclic structures visually. %Visual inspection relies heavily on these properties. 

\section{Experiments}
\label{Experiments}

Here, we illustrate the graphical power of PTA via Kuhn poker, RPS+2, Blotto, and Pokemon. All four exhibit interesting cyclic structures that are well explained by PTA. Kuhn is chosen to clearly illustrate the interpretation of each disc game as a trade-off, RPS+2 to show that the disc games can encode more complex trade-offs through the embedded geometry, Blotto to illustrate potential limitations of PTA arising from symmetries as well as the elegance of the trade-off representation when decomposing interlocked cyclic structures, and Pokemon to illustrate the diversity of possible trade-offs that can be recovered via PTA. In each case, we emphasize the interpretation of the principal trade-off to show that PTA reveals diverse, fine-grained game structures based only on empirical game data.  

%A simpler example is provided in Appendix \ref{sec: rps} for reference. 

\subsection{Kuhn Poker}

\begin{figure}[t]
    \centering
    \includegraphics[scale = 0.25]{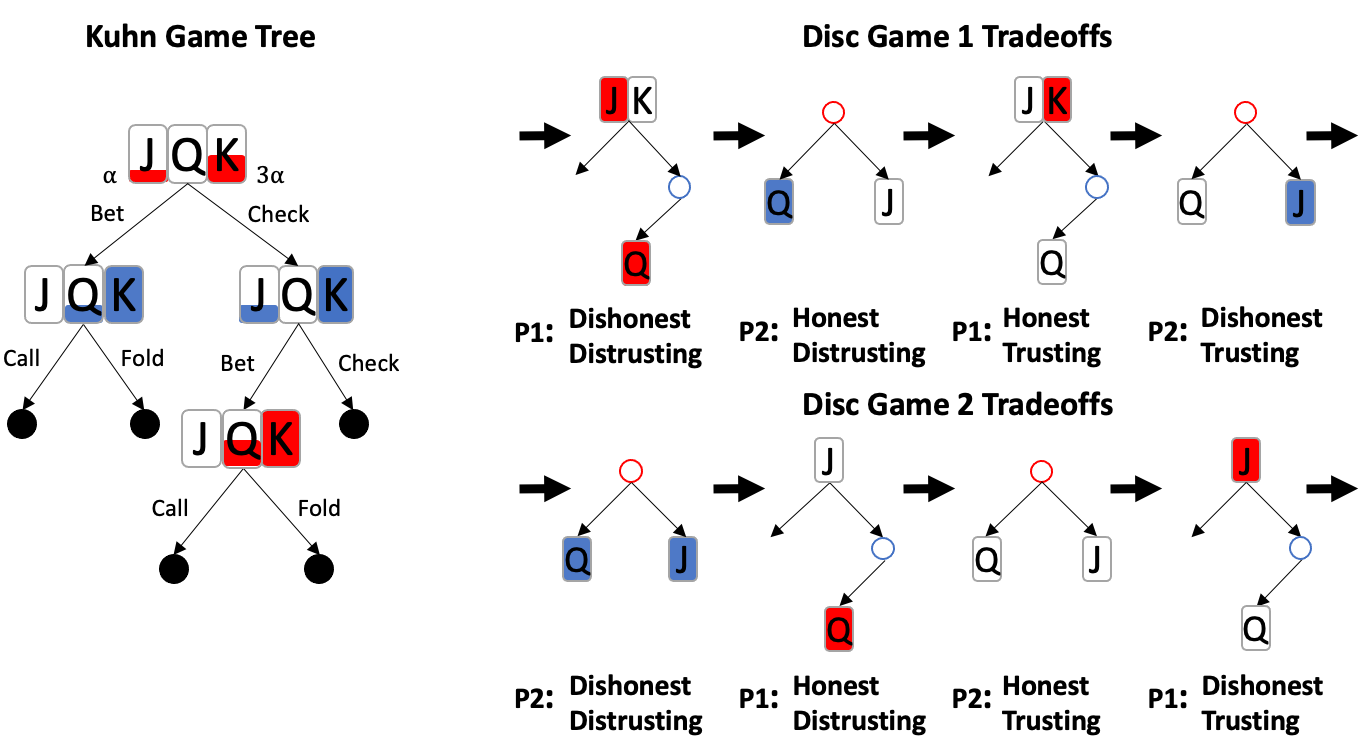}
    \caption{\textbf{Left:} The full 3 card Kuhn game tree. The height of each colored bar in a card indicates the probability a player with that card takes the more aggressive action. The frequencies shown are a Nash equilibrium for the Bet / Call actions. For example, at the first node, a player holding a queen should always check. Black circles are leaf nodes. \textbf{Right:} The tradeoffs in each disc game come from a subset of the game tree. Arrows point in the direction of advantage. Strategies that lead to tradeoffs are given labels below each reduced tree}
    \label{kuhn-tree}
\end{figure}

%% description of Kuhn poker
Kuhn poker is a simplified, two-player poker game played with an $n$ card deck (traditionally $n = 3$), single card hands, and a single betting round \cite{kuhn1950simplified}. It is also a zero-sum, incomplete information game.  Strategic trade-offs are produced by the lack of complete information. Each player's choices confer information about their hand to their opponent, however, that information may or may not be trustworthy. For example, a bet may or may not indicate a strong hand. 

The game tree for Kuhn poker is illustrated in Figure \ref{kuhn-tree} Given a particular pair of hands, the game tree includes 4 decision points. After a call, both hands are revealed, and the player with the higher card wins. Using an $n$ card deck, the extensive form game has $9 n (n-1)$ states and $2n$ information sets for each player. All choices are binary, and a particular agent could play as P1 or P2, so a complete policy consists of $4n$ decision probabilities. 

Given two agents, we measure the advantage one possesses over the other as their expected winnings if they are equally likely to bet first or second. Since no information set ever appears twice along the same path through the game tree, and no player ever makes more than two decisions per game, the payout function is a cubic, multilinear, skew-symmetric polynomial in the policies. 

The Nash Equilibrium policies (NE) for the three-card game are shown in Figure \ref{kuhn-tree}. Note that, at the NE, some decision probabilities are set to zero or one. These correspond to decisions for which only one action is rational. For example, it is never rational to fold when holding the strongest card or to call when holding the weakest. These decision probabilities lie at the boundary of the simplex of allowed policies and are subject to strong selection. We call the boundary containing the NE the rational boundary. 

Restricted to the rational boundary, the performance function $f$ is a bilinear function in the unconstrained decision probabilities. The game does not admit a unique NE, parameterized by the probability, $\alpha$, that the first player bets when holding a Jack (the weakest card) \cite{kuhn1950simplified}. Restricted to the boundary, the game is highly cyclic, since the unconstrained probabilities induce trade-offs associated with the information conveyed, and inferred, when betting or checking. These dynamics are entirely neutral when averaged over the rational boundary, thus restricted to the boundary, the game is favorite-free \cite{HHD}. That is, all policies on the rational boundary have zero expected payout when opponents are drawn from an isovariant distribution on the unconstrained policies.  

Nevertheless, policies lying off the NE on the rational boundary may be exploited by other strategies on the rational boundary. For example, when trained under optimal self-play, policies on the rational boundary circulate about the NE. These dynamics are directed by the self-play gradient vector field, $v(z) = \nabla_x f(x,y)|_{x=y=z}$. Note that a policy $z_*$ is only a NE if $v(z_*) = 0$. The circulation in $v(z)$ is driven by the aforementioned decision trade-offs. A dishonest player who tends to bluff (bet on a weak card) can be countered by a distrusting opponent who assumes a bluff. Assuming a bluff has zero net payout since it is countered by honest opponents but counters dishonest opponents. 

\begin{figure}[t]
    \centering
    \includegraphics[width=\textwidth]{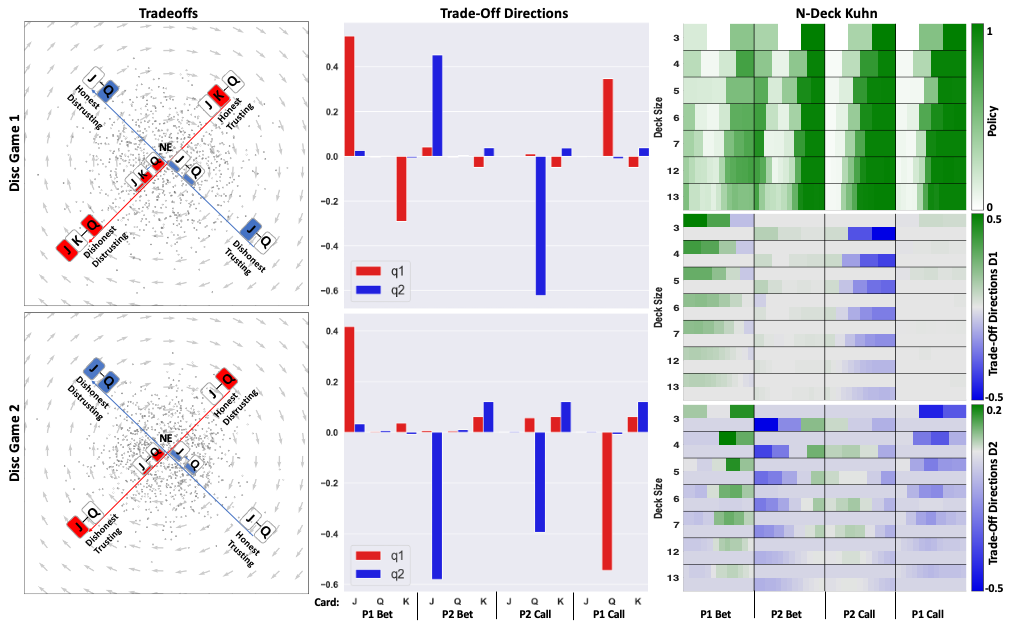}
    \caption{\textbf{Left:} The first two disc games with 4 marked axes representing the associated strategic trade-off about the NE. \textbf{Middle:} The weighting of the linear fit from policy to embedding is shown for each axis of the disc game. The $x$-axis labels shown at the bottom indicate the order of the weights w.r.t the policy. \textbf{Right:} An analysis of $N$-Deck Kuhn is visualized. The first table shows the Nash frequencies for various deck sizes. The next two tables show trade-off directions, as in the middle column for deck sizes ranging from 3 to 13. The color of each cell represents a decision probability (NE table), or adjustment to a NE probability. The consistency of the color patterns across deck size shows that the trade-off directions generalize over deck size.}
    \label{kuhn-discgames}
\end{figure}

%% description of experimental procedure
These considerations recommend Kuhn poker for PTA. We aim to recover, and rank by importance, the trade-offs that arise on the rational boundary using PTA. Unlike extant work, we target visualization and strategic feature extraction. 

%To recover the underlying trade-offs via PTA, we simulate a generic data generation process. PTA investigates game structure through its realization in a particular population. Thus, we require a relevant population to study.

To generate agents we used the Neural fictitious self-play (NFSP) algorithm \cite{heinrich2016deep} implemented using the Open Spiel software package \cite{LanctotEtAl2019OpenSpiel}. We used a simple feed-forward neural network with a single hidden layer of size 64 for both the policy and critic networks. We ran the algorithm until convergence, minimizing exploitability --- the best your opponent could do if they knew your policy. Near the NE, the exploitability will tend toward 0. To extract a policy representation we compute the decision probabilities from the converged policy network.

Given the converged agent, we identify particular policies that lie on or very near to the boundary of the allowed policy space. If the corresponding component of the self-play gradient points outward against the boundary, then we treat that policy as constrained to the boundary. Then, to generate a relevant population, we sample an ensemble of random agents near the converged policy, constrained to the rational boundary. Next, we evaluate performance between all pairs of agents in the population and apply PTA to embed them.

To extract the underlying trade-offs, we seek an interpretable mapping between policies and embedding coordinates. Since 3 card Kuhn poker is bilinear once constrained to the rational boundary, the map from policy to embedding coordinates is linear. For larger deck sizes, the performance function can be approximated by a quadratic polynomial provided the sampled agents are sufficiently close to the converged agent. Then, as in the 3 card case, the realized embeddings are near to linear functions of the agent policies \cite{strang2022quad}.

We recover that map by fitting the embedded agent coordinates to a linear function of their policies. The linear map into each disc game is a scaled projection onto a hyperplane in policy space. We identify the trade-offs by isolating the sparsest possible basis vectors for each hyperplane. 

We repeated this numerical experiment for decks of sizes 3 to 13. 

\begin{table}
  \centering
  \caption{Disc Game Importance in Kuhn Poker}
  \begin{tabular}{cccccccccc}
    \toprule
    Deck Size     & D.G. 1 & D.G. 2  & D.G. 3 & D.G. 4 & Total &  Error & (Num.~Rank)/2 & Game States & Info States  \\
    \midrule
    % ROW 1
    3 & 49\% & 30\% & 12\% & 8\% & 99\% & 1\%  & 3 & 54 & 12 \\
    % ROW 2
    4 & 50\% & 28\% & 12\% & 6\% & 96\% & 4\% & 3 & 108 & 16 \\
    % ROW 3
    5 & 50\% & 23\% & 13\% & 8\% & 94\% & 6\% & 4 & 180 & 20 \\
    % ROW 4
    6 & 48\% & 26\% & 11\% & 7\% & 93\% & 7\% & 4 & 270 & 24 \\
    % ROW 5
    7 & 49\% & 25\% & 11\% & 7\% & 92\% & 8\% & 4 & 378 & 28 \\
    % ROW 6
    13 & 46\% & 25\% & 10\% & 8\% & 89\%  & 11\% & 5 & 1,404 & 52\\
    
    \bottomrule
  \end{tabular}
  \vspace{2mm}
  \captionsetup{labelformat=empty}
 \caption{Importance of each disc game, measured as $\omega_k^2/(\sum_{k} \omega_k^2)$, for the first four disc games given populations concentrated near the NE in Kuhn poker with varying deck sizes.  The first column shows the deck size. The next six show the importance of the first four disc games, their total, and the associated reconstruction error in the rank $8$ approximation to $F$. The final three compare the number of disc games required to satisfy a 90\% accuracy threshold with the number of states in the game tree, and the dimension of the policy space.}
\label{tab: importances}
\end{table}

Even though the game tree grows quadratically in $n$, and the dimension of the policy space grows linearly in $n$, the number of relevant disc games (numerical rank of the performance matrix) remained close to constant across all deck sizes tested. To measure the numerical rank we computed the relative error in the low-rank approximation to $F$ for varying $r$. In all cases, the first two disc games accounted for at least 71\% of the structure of $F$, and the first four accounted for 89\% of the structure of $F$. The contributions of each disc game are reported in Table \ref{tab: importances}. Note that each disc game retains roughly the same importance, no matter the deck size. This observation suggests that the disc games recover strategic trade-offs that generalize over different deck sizes.

%% result summary
Figure \ref{kuhn-discgames} illustrates the trade-offs produced by the first two disc games when $n = 3$. These represent two different four-way rock-paper-scissor trade-offs. Figure \ref{kuhn-discgames} shows extremal policies that characterize the trade-offs. Note that these trade-offs arise directly from the incomplete information nature of the game. In each case, a player possesses an advantage over their opponent if they can correctly infer the opponent's hand based on their action, or can trick their opponent. The eight possible combinations of player role, honesty in action, and assumed honesty of opponent produce eight stereotypical policies, arranged in two disc games. The two disc games account for 49\% and 30\% of the observed performance relations. The first trade-off discovered in the 3-card game generalizes to larger card games (see the rightmost panels of Figure \ref{kuhn-discgames}). Thus PTA successfully extracts interpretable, generalizable strategic trade-offs from sample tournament data.

\section{Rock Paper Scissors + 2} \label{sec: rps}

Next, consider an extended example of the popular rock-paper-scissors (RPS) game, chosen to show that the geometry of the point cloud formed by the embedded agents can embody the structure of a game. Unlike the Kuhn example, we do not focus on a direct analysis of the mapping from strategy space to disc game space. Instead we aim to show that the point clouds' shape naturally represents the game. We consider rock-paper-scissor-lizard-spock (RPS+2). The utility matrix for (RPS+2) is the circulant matrix generated by the row $[0,-1,1,-1,1]$ (rock $\succ$ paper and lizard, scissors and spock $\succ$ rock).
%
%\begin{equation}
%    U_{RPS+2} = \begin{bmatrix}
%    \phantom{-}0 & -1 & \phantom{-}1 & -1 & \phantom{-}1 \\
%    \phantom{-}1 &  \phantom{-}0 &-1 &  \phantom{-}1 &-1 \\
%   -1 &  \phantom{-}1 & \phantom{-}0 & -1 & \phantom{-}1 \\
%    \phantom{-}1 & -1 & \phantom{-}1 &  \phantom{-}0 &-1 \\
%   -1 &  \phantom{-}1 &-1 &  \phantom{-}1 & \phantom{-}0 \\
%    \end{bmatrix}
    %\label{rps_2_utility}
%\end{equation}

First, we generate a population of agents using fictitious self play (FSP). Each agent in the population is defined by a vector of length 5 representing a mixed strategy. We start with an initial random agent and generate best response agents using FSP. Each best response agent is added to the population. We then create $F$ by computing the expected value of each match-up using the utility matrix. 

\begin{figure}[t]
    \centering
    \includegraphics[width=\textwidth]{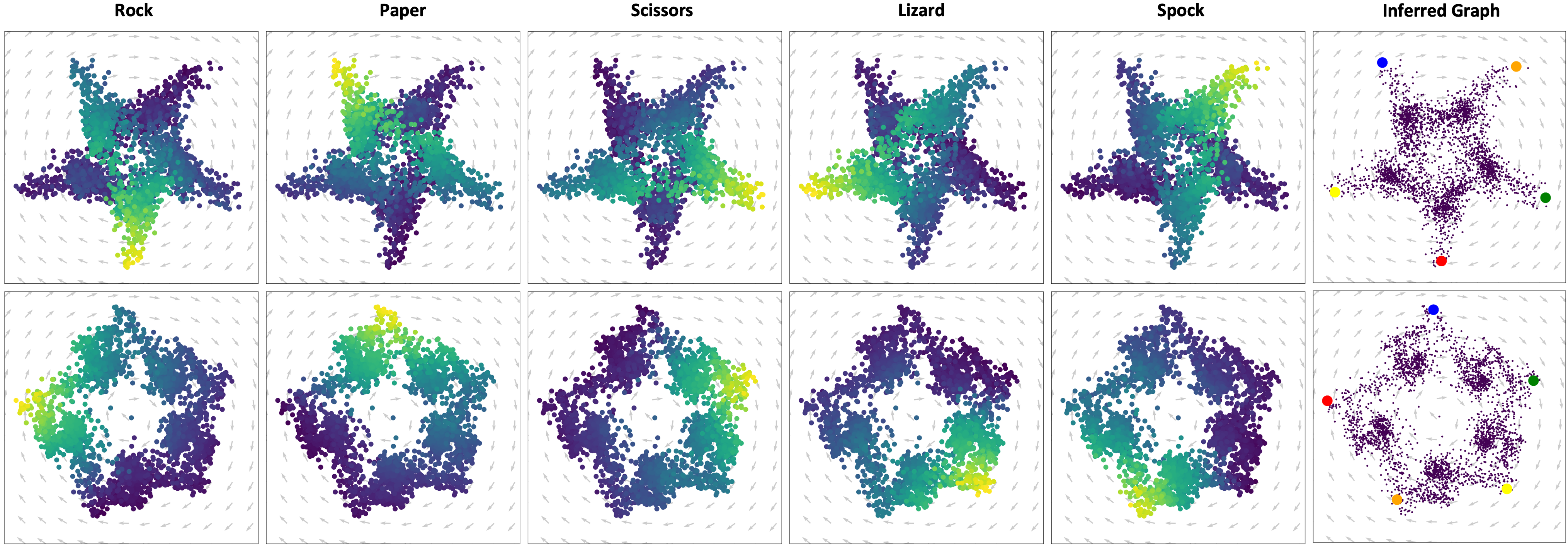}
    \caption{The two disc games that represent RPS+2 are shown. Starting from the first column the points are colored by the label for that column. The graph on the left shows a reduced representation of the game by using representative points for each strategy (red for rock, blue for paper, green for scissors, yellow for lizard, orange for spock).}
    \label{rps-2}
\end{figure}

What embedding should we expect? All mixed strategies are an interpolation of the 5 pure strategies, and the game is bilinear in the action probabilities, so the embedding map must be approximately linear. Thus, each mixed strategy should be contained inside the convex hull of a polygon formed by embedding the pure strategies. The utility matrix is invariant under cyclic permutations, so the polygon must be regular and centered at the origin. Therefore, the pure strategies must be the vertices of a regular, centered, pentagon. 

The full game includes two, equally important, interlocking cyclic relations among the pure strategies. These are, rock $\succ$ paper $\succ$ scissors $\succ$ lizard $\succ$ spock $\succ$ rock, and rock $\succ$ lizard $\succ$ paper $\succ$ spock $\succ$ scissors $\succ$ rock, where $\succ$ denotes loses to. Both cycles contribute equally to the utility matrix as they are equivalent under a relabeling. Namely, the utility matrix is unchanged under a permutation that rearranges the strategy labels so that rock is followed by lizard, then paper, then spock, then scissors.

These two cycles are represented by a pair of disc games, as shown in Figure \ref{rps-2}. In both, the pure strategies are vertices of a regular pentagon centered at the origin. The ``Inferred Graph" in the rightmost column plots a representative point for each of the pure strategies. Advantage relations that are not accounted for in the first disc game are represented in the second. In the first disc game, the sequence of pure strategies, Rock, Paper, Scissor, Lizard, Spock, are spaced by $144^{\circ}$ degrees about the pentagon. In the second, the sequential pure strategies are spaced $72^{\circ}$ apart, so occupy adjacent corners of the pentagon. The sequence of vertices produced by skipping $144^{\circ}$ per step, or moving $72^{\circ}$ degrees per step, are related by the same permutation that exchanged the two interlocked advantage cycles defined by the utility matrix. The star shaped disc game has a larger eigenvalue, since $|\sin(144^{\circ})| < |\sin(72^{\circ})|$ so vertices of the star must be moved farther from the origin to maintain the same cross product as adjacent vertices of the pentagon in the second disc game.  

As in Kuhn poker, the patterns uncovered by PTA repeat for larger RPS games, whose embedded point clouds form regular, centered polygons with as many vertices as pure strategies, as many disc games as cycles in the utility matrix, and whose vertex orders are related by the permutations which exchange cycles in the utility matrix.

%\begin{figure}[t]
 %  \centering
 %   \includegraphics[width=\textwidth]{Figures/rps_game_graph.png}
%    \caption{The first row shows the abstracted performance graphs for the full game as well as each disc game. Pure strategies are represented by color (red for rock, blue for paper, green for scissors, yellow for lizard, orange for spock). Arrows represent dominance relations. For example, the arrow from green to red in the middle column denotes the strong advantage rock possesses over scissors due to the first disc game. The heat maps in the second row represent performance matrices for the full game and for each disc game. The full performance relation is a sum of the two advantage cycles produced by each disc game.}
    \label{rps-2-game-graph}
%\end{figure}

%To clarify the advantage relations encoded by each disc game, we use the inferred graph to construct an abstract representation of a performance graph for the pure strategies. The result is shown in Figure \ref{rps-2-game-graph}. Colors denote strategy, and arrows denote advantage. The full game graph can be decomposed into the two separate game graphs embedded in each of the disc games.  

%We also show heatmaps of performance for each of the disc games. Here performance is represented as row vs column agent. The full performance matrix $F$ can be fully decomposed into a sum of disc games as described in \ref{PTA}. The matrix for disc game 1 has entries where the performance of the row vs column player can always be described as an interpolation of the game graph shown in the row above. The same goes for disc game 2. This example serves as a demonstration for what PTA is doing. $F$ was decomposed into a series of disc games where each disc game represented unique trade-offs. Similarly as with Pokemon in the main text we could have approximated the utility matrix \ref{rps_2_utility} of the game just from observing the performance graphs of each disc game. 

\subsection{Colonel Blotto}

\begin{figure}[t]
    \begin{center}
        \centerline{\includegraphics[width=\textwidth]{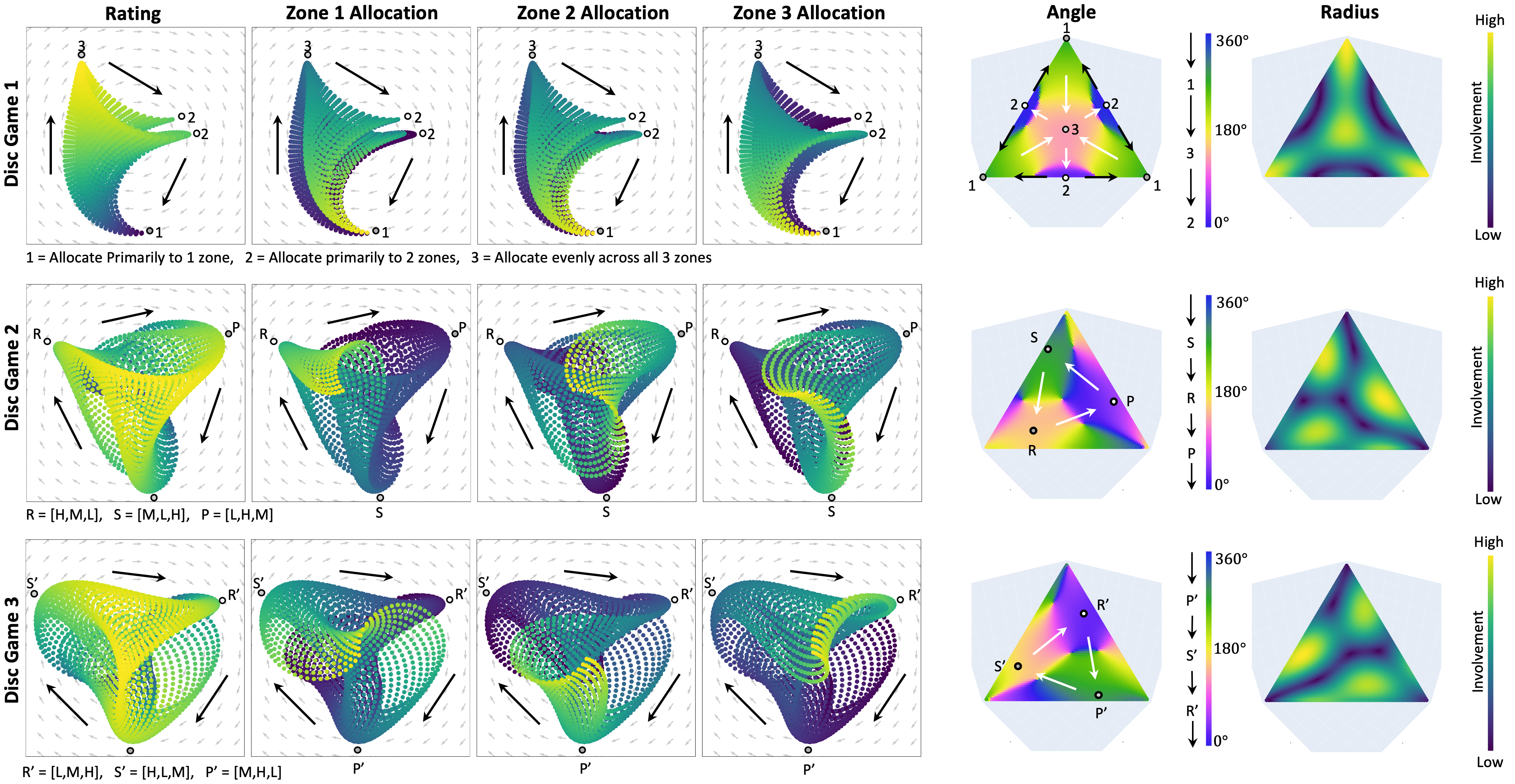}}
    \end{center}
    \caption{Disc games 1 to 3 of Blotto [1,1,1] game with $N$ = 75. \textbf{Rows:} disc game number. \textbf{Columns 1 - 4:} disc game embeddings colored according to agent rating, then agent allocation to zones 1 to 3. \textbf{Column 5:} the angle (measured counterclockwise from the horizontal axis), of the embedding of each strategy. Advantage flows clockwise in angle, so blue beats purple beats yellow beats green beats blue. \textbf{Column 6:} the radius of the embedding of each strategy. High radius corresponds to strong involvement in a trade-off (yellow), while small radius corresponds to low involvement (blue). Each triangle in the fifth and sixth columns represents the space of available allocations. High allocations to zone 1 cluster near the bottom left corner, high allocations to zone 2 cluster near the bottom right corner, and high allocations to zone three cluster at the top corner. \textbf{Labels:} Representative allocations defining the underlying trade-offs, indicated with bold arrows (see Table \ref{tab: tradeoffs}).}
    \label{figure-2.1}
\end{figure}

Colonel Blotto is a zero-sum, simultaneous action, two-player resource allocation game \cite{Blotto-Generalizations}. Each player possesses $N$ troops to distribute across $K$ zones. Each zone has an associated payout $Z_k$. A zone is conquered by a player if they allocate more troops to that zone than their opponent. The conquering player receives the payout. Ties result in both players receiving 0 payout. The player with the highest total payout wins the match. All allocations are revealed simultaneously.

At simplest, the payouts are uniform across zones, so the player who conquers the most zones wins the game. Unweighted Blotto is a highly cyclic game since there is no dominating strategy. Every strategy admits a counter. Unless $K = N$ or $K <= 2$, all allotments lose to some other allotment. To defeat an allotment, adopt the maxim, ``lose big, win small". Mimic the allotment, then redistribute all the units from the zone with the most units as uniformly as possible across the remaining zones. Then, unless all zones were allotted one unit, the exploiting strategy sacrifices a loss in one zone to win in more than one other zone. In general, the more an allotment commits to a single zone, the more easily it is defeated. Unweighted Blotto is also complex, since the zones are indistinguishable. Thus unweighted Blotto admits $K!$ fold symmetries with respect to the zone labels. 

%Introducing nonuniform weights changes the game by breaking the exchange symmetry of the zones. This changes the set of possible win conditions, which has direct implications for the overall cyclicity, complexity, and strategic trade-off structures revealed by PTA. 

We consider each unique strategy as a separate ``agent'', parameterized by the corresponding allotment. We generate agents by randomly sampling over the strategy space using a Dirichlet distribution with the support equal to the number of zones. %This method is tractable for the smaller games we consider. 
After sampling, we compare each pair of strategies in the population. Each match-up is deterministic and results in a win, loss or tie, which we assign scores (0.5,0,-0.5). We construct the associated evaluation matrix by setting $F_{ij}$ to the score of strategy $i$ against strategy $j$.

%\textbf{Blotto Example 1}

% Substituting in Booktabs table
\begin{table}
  \centering
  \caption{Principal Trade-Offs}
  \begin{tabular}{llllc}
    \toprule
    %\multicolumn{5}{c}{Principal Trade-Offs} \\
    \cmidrule(r){1-5}
    D.G.     & Allocation Types & Location in Simplex  & Example & Advantage Relation \\
    \midrule
    % ROW 1
     \vspace{2mm}
      1
      & \begin{tabular}{@{}l@{}l@{}} 
          $(1) = \text{allocate to 1 zone}$  \\ 
          $(2) = \text{allocate to 2 zones}$ \\  
          $(3) = \text{allocate to 3 zones}$
        \end{tabular} 
      & \begin{tabular}{@{}l@{}l@{}} 
          corners  \\ 
          midpoints of edges \\  
          center
        \end{tabular} 
      & \begin{tabular}{@{}l@{}l@{}} 
            $[\textbf{70},0,5]$ \\ 
            $[\textbf{38},\textbf{37},0]$ \\  
            $[\textbf{25},\textbf{25},\textbf{25}]$
        \end{tabular} 
      & $(1)$ < $(3)$ < $(2)$ < $(1)$ \\
    % ROW 2
      \vspace{2mm}
      2 
      & \begin{tabular}{@{}l@{}l@{}} 
          $R = [H,M,L]$  \\ 
          $S = [M,L,H] $ \\  
          $P = [L,H,M]$
        \end{tabular} 
      & \begin{tabular}{@{}l@{}} 
          corners  \\ 
          shifted \\  
          counter clockwise
        \end{tabular} 
      & \begin{tabular}{@{}l@{}l@{}} 
          $ [\textbf{50},25,0]$ \\ 
          $ [25,0,\textbf{50}]$ \\  
          $ [0,\textbf{50},25]$
        \end{tabular} 
      & $R$ < $P$ < $S$ < $R$ \\
    % ROW 3
      3 
      & \begin{tabular}{@{}l@{}l@{}} 
          $R' = [L,M,H]$  \\ 
          $S' = [H,L,M] $ \\  
          $P' = [M,H,L]$
        \end{tabular} 
      & \begin{tabular}{@{}l@{}} 
          corners  \\ 
          shifted \\  
          clockwise
        \end{tabular} 
      & \begin{tabular}{@{}l@{}l@{}} 
          $ [0,25,\textbf{50}]$ \\ 
          $ [\textbf{50},0,25]$ \\  
          $ [25,\textbf{50},0]$
        \end{tabular} 
      &  $R'$ < $P'$ < $S'$ < $R'$ \\
    \bottomrule
  \end{tabular}
  \vspace{2mm}
  \captionsetup{labelformat=empty}
 \caption{Principal trade-offs associated with the first three disc games. The columns list the allocation types involved in each disc game (D.G.), their location in the simplex of possible allocations, provide an example set of allocations, and the competitive relations between the types. The letters H, M, and L, are used to denote high, medium, and low allocation. Note that the allocations involved in the trade-off defined by a disc game correspond to locations in the radius panels in Figure \ref{figure-2.1} shaded green or yellow. Advantage in a disc game flows clockwise, so can be inferred from the angle panel. }
\label{tab: tradeoffs}
\end{table}

PTA allows elegant visualization of relevant game structure by reducing a game to a small set of key trade-offs. We start by looking at the $K$ = 3, $N$ = 75 blotto game with uniform payouts. Table \ref{tab: tradeoffs} summarizes the principal trade-offs associated with each disc game. These trade-offs are the most important sources of cycles in the tournament, accounting for 80\% of its structure.

In general, the number of distinct allotments in a $K$ battlefield, $N$ troop blotto game grows at $\mathcal{O}(N^K)$, but the complexity, which reflects the underlying number of cyclic modes, converges to a constant value associated with a continuous Blotto game, where commanders can allocate an arbitrary fraction of their force to each zone. Unweighted $K$ = 3, $N$ = 75 blotto admits 2,926 allotments, but has a 3! fold exchange symmetry under permutations of the battlefield labels, leaving roughly 488 distinct allotments. Three disc games reconstruct the evaluation matrix to $\approx$ 80\% accuracy, 6 to $\approx$ 90\% accuracy, and 12 to $\geq$ 95\% accuracy, so the game has complexity 12 at a 95\% standard. Trade-offs 4 - 12 represent refinements of the trade-offs present in the first three disc games, so PTA really allows a reduction in complexity from 2,926 allocations (absent prior knowledge regarding symmetries), to 3 fundamental cyclic modes. Thus, PTA can effectively separate the underlying complexity of a game from the size of its strategy space.

%% paragraph on symmetries: repeated eigenvalues... can observe symmetric structure in the eigenvalues, induces degeneracy in the representation, really six dimensional objects, need to consider multiple disc game at once. In this case, still revealing
The exchange symmetry of the zones is apparent in the sequence of eigenvalues, $\omega_k$, representing disc game importance. Exchanges introduce 6 permutations under which the evaluation matrix is invariant. Consequently, $\omega_k$ come in sets of three, where each $\omega_k$ represents a pair of eigenvectors. Eigenvectors associated with identical eigenvalues are not uniquely defined. Instead, they are drawn from a subspace of dimension equal to the multiplicity of the repeated eigenvalue. Consequently, all of the eigenvectors $Q$ are chosen arbitrarily from six dimensional spaces.

When $F$ has repeated eigenvalues, the associated disc game embeddings are not unique. Any unitary transform of the set of eigenvectors sharing an eigenvalue defines a valid embedding. Thus, symmetry presents an unusual challenge: degeneracy. In our case, the disc games come in sets of three, each representing an arbitrary rotation of a six dimensional object. Consequently, we consider multiple disc games simultaneously. This issue was not addressed in previous work, which largely only considered the leading disc game. Generic games should not exhibit strong symmetries, so such degeneracy will be rare and confined to toy examples. That said, generic games also require more than one disc game, so it is essential to consider more than the leading disc game.

%% disc game analysis
We analyze the three leading disc games to identify the most important allocation trade-offs. Figure \ref{figure-2.1} shows the first three disc games colored by rating, allocation to the three zones, and the mapping to angle and radius in each disc game as a function of allocation. Each share the same eigenvalue, so are equally important and could be mixed. Nevertheless, these three disc games represent distinct trade-offs in allocations that can be easily explained. 

The specific trade-offs can be identified directly from the disc games when colored by allocation. Consider the points labeled 1, 2, and 3 in the first disc game. Each maximize the radius of the scatter cloud while moving along its boundary, so represent the allocations most involved in the cycle. The low rated points at the bottom of the scatter allocate primarily to one zone (yellow in panels 2 - 4). 

Moving clockwise, the next extrema occurs at the top of the scatter. It is high rated, and has nearly equal allocation across all three zones (colored green in panels 2 - 4). Uniform allocations are rated highly since they perform well against most randomly sampled allocations, particularly those lying along a line connecting a corner of the simplex to its center. This induces a transitive trend among the bulk of the allocations moving from allocations that prioritize one zone, to allocations that treat the zones equally. This transitive trend is represented by the general shift of the disc game leftward off the origin. This subset of allocations compete transitively, producing the regular gradient from purple to yellow in rating when moving clockwise from the bottom to the top in the scatter.

Not all allocations satisfy this transitive trend. Allocations that prioritize two zones counter the uniform strategy, and are countered by allocations that prioritize a single zone. For example, allocation [70,0,5] defeats [38,37,0]. Thus, allocations lying on the midpoints of an edge of the simplex lose to allocations near either neighboring endpoint. These counters close the cycle, and are represented by the rightmost pair of corners labelled 2 in disc game 1. Panels 2 - 4 show that each such corner receives an intermediate allocation in two zones (green), but little to none in the third (dark blue). 

\begin{figure*}[t]
    \begin{center}
        \centerline{\includegraphics[width=\textwidth]{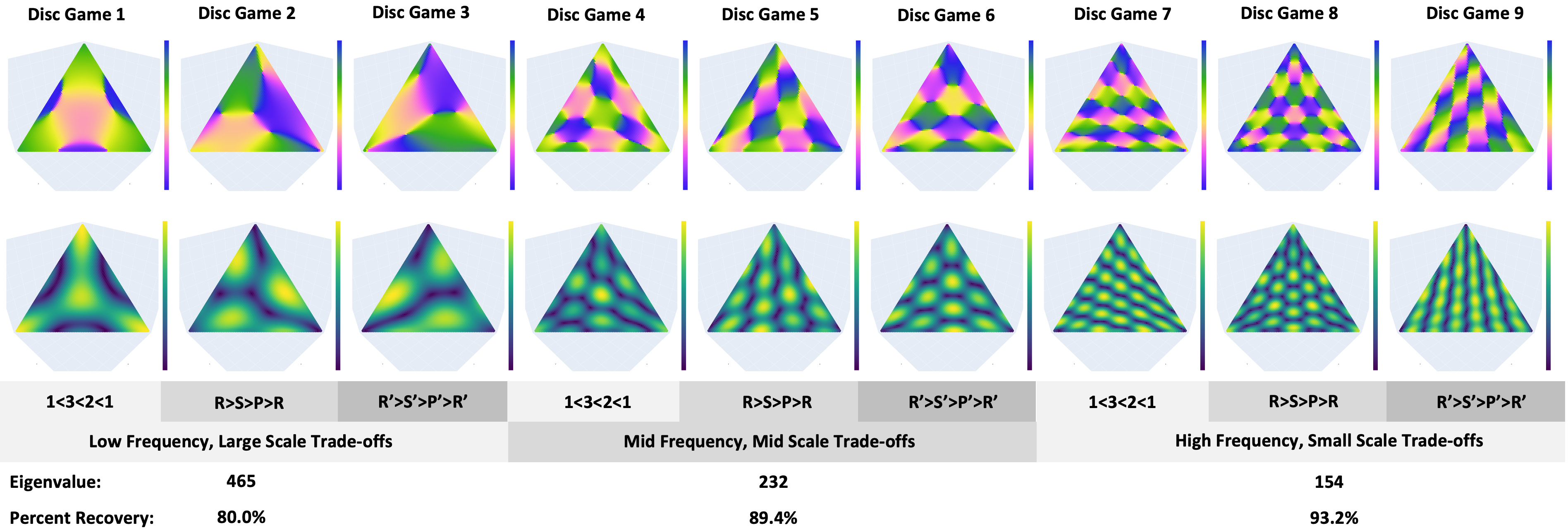}}
    \end{center}
    \caption{The first nine disc games of the $N$ = 75, [1,1,1] blotto game. Each column is a separate disc game. The first and second rows show the angle and radius assigned to each allocation. The disc games are labelled by trade-off type. Consecutive sets of three share the same eigenvalue and are grouped by spatial scale with eigenvalue and percent recovery of $F$ provided beneath.}
     \label{fig: [1,1,1] games 1 to 9}
\end{figure*}

Similar visual analysis identifies the RPS cycles among cyclic permutations of allocations [H,M,L] and [L,M,H] shown in disc games 2 and 3. For example, the leftmost corner of the scatter cloud shown in disc game 2 receives a high allocation in zone 1 (teal), an intermediate allocation in zone 2 (blue-green), and a low allocation in zone 3 (dark blue). Walking from R to P to S, the allocation patterns shifts cyclically. The same analysis applies to disc game 3, starting from [L,M,H].

Figure \ref{fig: [1,1,1] games 1 to 9} shows the angle and radius assigned to each allocation in the simplex. Strikingly, subsequent disc games imitate the disc game 1-3 trade-offs, only at higher frequency on a smaller spatial scale in allocation. This suggests that the disc games may act like Fourier modes, where early disc games capture low frequency, global trade-offs, and later disc games capture high frequency, local trade-offs. It also suggests that orthogonality may not be the appropriate notion of independence for trade-offs. A sharper notion of equivalency is needed. Methods like nonnegative matrix factorization, which address similar issues among PCA features \cite{lee1999learning}, suggest an avenue for further development. An example that produces explicit sine series is discussed in the Appendix. 

\subsection{Pokemon}

 We conclude by analyzing Pokemon. Pokemon originated from the Nintendo Game Boy console, but has since been played on a variety of mediums including playing cards. %\cite{pokemon-site}. 
Pokemon is of considerable interest from a game design perspective since the creators must design certain trade-offs to keep the game balanced and engaging. The game is made up of creatures, called Pokemon, that come in many varieties. Players are rewarded for collecting diverse teams. Thus, each Pokemon has a different type, and each type has its own set of strengths and weaknesses. These different types satisfy interlocking cyclic relationships.

The data used in this analysis comes from an open-source Kaggle data set \cite{pokemon-data}. The original data has 800 Pokemon, but we removed the 65 ``legendary" Pokemon to simplify the analysis. The data consists of battle outcomes and pokemon attributes. Battle outcomes were converted into an evaluation matrix by logistic regression (see Appendix \ref{appendix:pokemon}). %Here, we apply the Schur decomposition directly to $F$ to show that disc game embedding can successfully isolate a dominant transitive component.

\begin{figure*}[t]
    \begin{center}
        \centerline{\includegraphics[width=\textwidth]{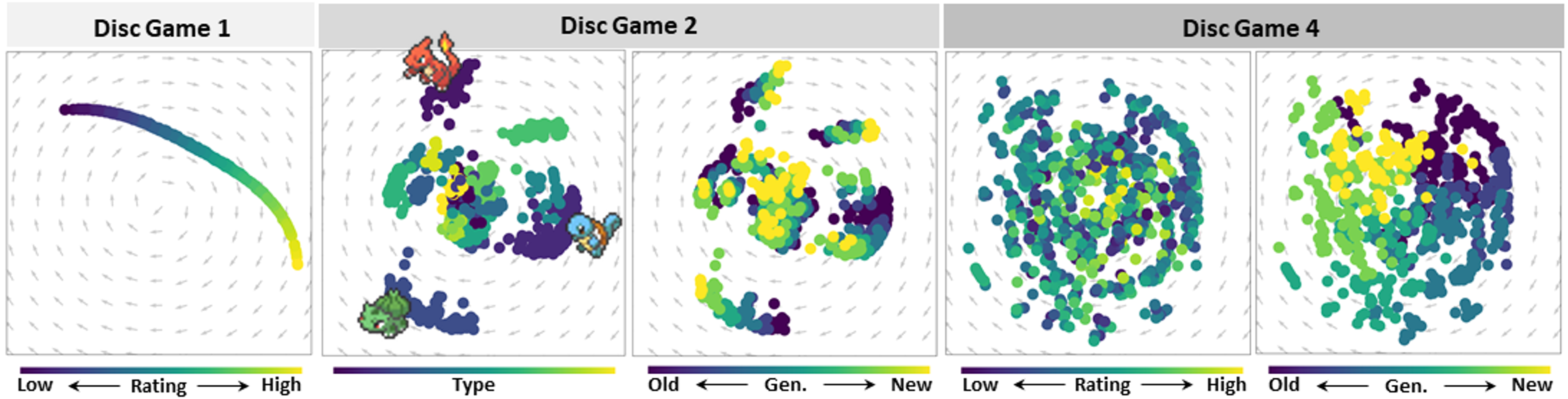}}
    \end{center}
    \caption{Disc games 1,2 and 4 for pokemon. Disc game one is colored by rating. Disc game two is colored by type, then generation. Disc game four is colored by rating, then generation. }
    \label{pokemon-disc-games}
\end{figure*}

Figure~\ref{pokemon-disc-games} shows three of the first four disc games, chosen for their significance. The first disc game is the most important, and is clearly transitive since all points fall on a curve that does not enclose the origin. Position along the curve is closely correlated with speed, so speed determines rating. %Speed controls which Pokemon goes first in a battle, so strongly predicts average performance.  

\begin{figure}[t]
    \includegraphics[width=\textwidth,scale=0.9]{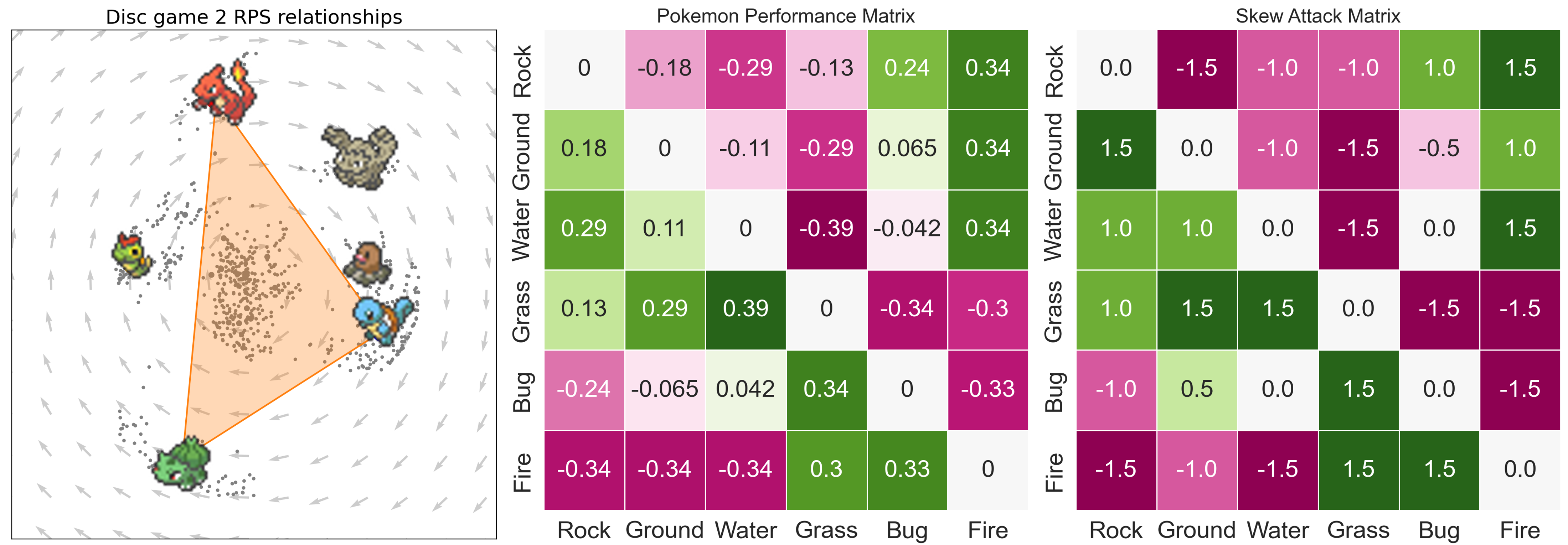}]
    \caption{\textbf{Left:} RPS sub-game discovery. Each cluster type is represented by a matching pokemon, \textbf{Middle:} Empirical performance matrix, \textbf{Right:} Performance matrix derived from \cite{pokemon-attack-matrix}}
    \label{pokemon-panel3} 
\end{figure}

We query by attribute to interpret the remaining disc games. To start, consider the ``type" attribute. The second disc game is clearly clustered by type (see Figure~\ref{pokemon-disc-games}). A variety of RPS relationships are apparent among the type clusters. Any loop of clusters containing the origin corresponds to a cycle of type advantage. The intensity of the corresponding cycle (curl) is proportional to its area. Focus on the large clusters most involved in the trade-off, i.e. furthest from the origin. Figure~\ref{pokemon-panel3} summarizes the RPS relations between these clusters. First, notice the highlighted triangle formed by the Water-Fire-Grass clusters. The disc game shows the expected advantage cycle since the triangle contains the origin. Thus, PTA identifies known game structures without domain specific knowledge.

Additional clusters on the outer ring are more intricately related. The other three types are ``bug", ``rock" and ``ground". To summarize these relations we construct a coarse grained evaluation matrix, $\hat{F}$. Specifically $\hat{F}_{ij}$ is the average performance of Pokemon from type $i$ vs the Pokemon from type $j$ in the second disc game. The associated matrix heat-map is shown in the middle panel of Figure~\ref{pokemon-panel3}. The types are ordered by angle moving clockwise about the origin. 

We compared these relations with  available game design matrices known as ``attack matrices", which
list the advantage of one Pokemon type over the other. We use the attack matrix from (\cite{pokemon-attack-matrix}). An attack matrix is written in terms of multiples, so Pokemon that are evenly matched have a $1 \times$ advantage. We bucket the range of $i,j$ attack multipliers $a_{ij}$ into 5 bins ranging from $0 \times$ to $2 \times$, skew-symmetrize via $(A - A^{T})$. The result is the rightmost panel in Figure~\ref{pokemon-panel3}.

The coarse grained summary $\hat{F}$ is strikingly similar to the provided attack matrix. The apparent structural parity in these two matrices highlights the virtues of PTA. Without any domain knowledge, access to attributes, or any explicit instruction to identify clusters, PTA clustered Pokemon by their most relevant attributes (type) then encoded a game mechanism (type specific attack multipliers) directly from the cluster locations. Conversely, the second disc game shows how cyclic relations introduced at the mechanism level are realized in actual performance.

Coloring the disc games by ``generation", i.e.~pokemon release date, reveals design choices. The game is frequently updated by the addition of new Pokemon. Updates present a design challenge. Game designers must introduce desirable new Pokemon without upsetting the game balance. The fourth disc game, shown in the far right plot of  Figure~\ref{pokemon-disc-games}, is balanced in that rating does not predict angle, and instead correlates with radius. Strong and weak Pokemon are closer to the origin, while Pokemon of intermediate rating are more involved in the trade-off. This reveals a spinning top structure characteristic of many games \cite{czarnecki2020real}. Rather, generation predicts angle. Each generation possesses an advantage over its predecessor, as illustrated by the fade from purple to yellow. Balance is retained since generational advantage is periodic. The same clockwise generation shift reappears in the second disc game. Within type, new beats old. For example, the bottom-most cluster (grass) clearly trends old to young. Cross type relations are largely unchanged.

\section{Conclusion}

Following Balduzzi \cite{balduzzi2018re}, we have demonstrated that all evaluation matrices admit an expansion onto a sum of disc game embeddings. We suggest the name PTA based on the close analogy with PCA. Through examples, we have demonstrated that embeddings produced by PTA can reveal a surprising variety of competitive structures from outcome data alone. %Without prior knowledge of Pokemon, PTA was able to reveal trade-offs in the game arising from speed, type, and generation. Game design choices related to both type and generation were discovered absent prior knowledge, and then confirmed. Likewise, without any knowledge of the game rules, or win conditions, PTA identifies symmetries, and win condition trade-offs in Blotto. 
%These methods can be applied to any 2 player constant sum game, or any decision problem involving pairwise choice. 
Future work could provide more general methods for finding embeddings, such as a functional theory connecting performance with attribute space, or could seek a sparser representation via extensions of sparse PCA \cite{zou2006sparse}. Future work should also investigate automated methods that summarize the trade-offs identified by PTA without the need for visual inspection, and that leverage the representation for game classification, construction, and exploration.

\bibliographystyle{siam}
\bibliography{main.bib}

\appendix
\section{Appendix (Supplementary Material)}\label{Appendix}
\input{./appendix_for_Rev_3}

\end{document}

%% file: math_commands.tex
\usepackage{amsmath,amsfonts,bm}

\def\eqref#1{equation~\ref{#1}}
\def\1{\bm{1}}

\DeclareMathAlphabet{\mathsfit}{\encodingdefault}{\sfdefault}{m}{sl}
\SetMathAlphabet{\mathsfit}{bold}{\encodingdefault}{\sfdefault}{bx}{n}

%% file: appendix_for_Rev_3.tex
\maketitle

\section{Principal Trade-off Analysis}\label{appendix:pta}
% Put derivation of sum of disc games here

\subsection{Schur Decomposition is a Sum of Disc Games}

Here we prove that the Schur decomposition (real Schur form), is equivalent to a sum of disc games applied to the embedding maps $\vec{y}_k$. 

Recall the embedding construction. Given a skew symmetric matrix $F$, write $F = Q U Q^{T}$ where $Q$ is real, orthonormal, $U$ is block diagonal with diagonal blocks $\omega_k R$, and $R$ is the two by two ninety degree rotation matrix. Let $\vec{y}_k(i) = \omega_k^{1/2} [q_{i,2k-1},q_{i,2k}]$. Then, the rank $2r$ approximation to $F$ is:
\begin{equation}
 \begin{aligned} F_{ij}^{(2r)} & = e_i^{T}\left(\sum_{k=1}^{r}w_{k} [q_{2k -1};q_{2k}]^{T} R [q_{2k -1};q_{2k}] \right)e_j \\
& = \sum_{k=1}^{r}w_{k} [q_{i,2k -1};q_{i,2k}]^{T} R [q_{j,2k -1},q_{j,2k}] \\
& = \sum_{k=1}^{r}w_{k} (q_{i,2k -1} q_{j,2k} - q_{i,2k},q_{j,2k -1}) \\
& = \sum_{k=1}^{r}\sqrt{w_{k}} [q_{i,2k -1},q_{i,2k}] \times \sqrt{w_{k}} [q_{j,2k -1},q_{j,2k}] \end{aligned}
\end{equation}
Recalling the embedding construction, write:
\begin{equation}
F_{ij}^{(2r)} =  \sum_{k=1}^{r} \vec{y}_k(i) \times \vec{y}_k(j) = \sum_{k=1}^{r} \text{disc}(\vec{y}_k(i),\vec{y}_k(j)).
\end{equation}

Thus, restricted to each planar embedding $F_c^{(2r)}$ is a disc game and the optimal rank $2r$ approximation of $F^{(2r)}$ is a linear combination of disc games applied to the sequence of planar embeddings $\{\vec{y}_k\}_{k=1}^{r}$. $\square$

\subsection{PTA and Fourier Series}

%% note the apparent modal behavior
Both the $[1,1,1]$ and the $[2,3,4]$ blotto examples exhibit strikingly modal disc games that repeat at increasing frequency, and on smaller spatial scales, with increasing disc game number. These patterns suggest an analogy to Fourier series. To make the analogy more concrete we present one last example. 

%% show the [1,2,4] example + figure
Consider $[1,2,4]$ blotto. Since the net value of the first two zones is less than the value of the fourth zone, a player wins the overall game if they they win the third zone, or, tie in the third zone and win the second. Otherwise they tie in all zones or lose. Thus, the performance function $f([x_1,x_2,x_3],[y_1,y_2,y_3]) = \text{sign}(x_3 - y_3) + \chi_{z_3=0}(x-y) \text{sign}(x_2 - y_2)$ where $\chi(z)$ is the indicator function for the event in the subscript. Performance can be reduced to a comparison of a single agglomerated trait, $w(x) = x_3 + \frac{1}{N - x_3 + 1} x_2$. Then $f(x,y) = \text{sign}(w(x) - w(y))$. Thus, performance is a step function applied to the difference $w(x) - w(y)$. The difference is dominated by the difference in allocations to the third zone. 

\begin{figure}[t]
    \centering
    \includegraphics[width=\textwidth]{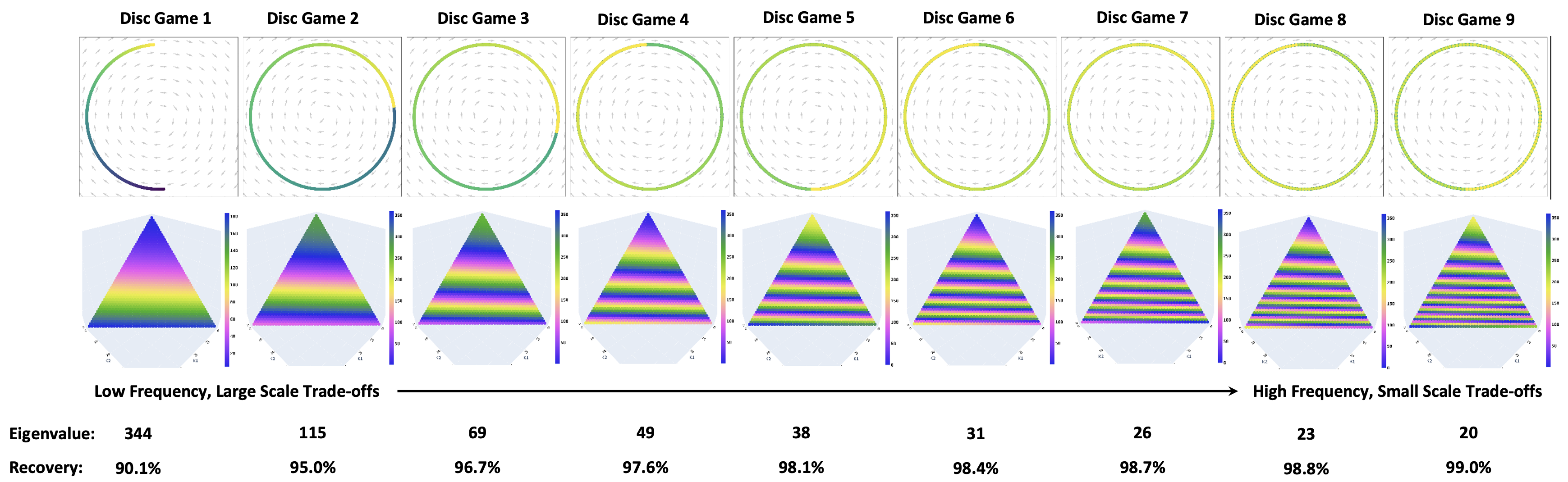}
    \caption{ \textbf{Top row:} The first 9 disc games for $[1,2,4]$ blotto, colored by rating. \textbf{Bottom row:} Phase as a function of allocation in each disc game. The top corner of the triangle corresponds to exclusive allocation to zone 3. Note that phase allocation increases in frequency with increasing disc game. Colors denote phase, with advantage flowing clockwise. The eigenvalues associated with each disc game are provided, along with the percent of $F$ recovered by the running sum of the disc games. Note that the first disc game is responsible for 90\% of the structure of $F$, and the first two disc games account for 95\% of its structure.}
    \label{fig: blotto fourier}
\end{figure}

Figure \ref{fig: blotto fourier} shows the first nine disc games. Notice that all allocations are embedded onto circles, or, in the first case, half circles. Moreover, phase (position along each circle), is entirely a function of the agglomerated trait $w(x) = x_3 + \frac{1}{N - x_3 + 1} x_2$. This form is apparent in the phase panels, where phase is close to constant for fixed allocation to zone 3, but is tilted slightly to account for allocation to zone 2. The first disc game is transitive and completes one half circle moving clockwise from zero allocation to zone 3 to exclusive allocation to zone 3. Discs games 2 and 3 complete 1.5 and 2.5 rotations each when moving from zero allocation to zone 3 to exclusive allocation to zone 4. The pattern continues for the first 9 disc games. Moreover, the circle radii decay geometrically. 

These features are hallmarks of a sine series embedding. We show below that any translationally invariant performance function of a single attribute can be represented by a sum of disc games, where the embedding into each disc game maps to a circle, the attribute maps to a phase coordinate around the circle, and the radii of the circles in each disc game are controlled by the coefficients of a sine series expansion. The performance function $f(x,y)$ only depends on the difference in allocations, $x - y$, so is translationally invariant, and is a function of a single trait, $w(x)$. Thus, it admits a sine series expansion. Note, the subsequent analysis does not guarantee low rank optimality, so only shows that disc game embedding via sine series is possible, not that PTA will necessarily produce such an embedding. 

%% show how to embed a sine function
Consider a performance function of the form:
\begin{equation} \label{eqn: sin performance}
    f(x,y) = A \sin(2 \pi \omega (x - y))
\end{equation}
for $x,y \in \Omega \subset \mathbb{R}$, for arbitrary amplitude $A$, and frequency $\omega$. 

Performance functions of this form are easy to embed, since the disc game uses a cross product. The cross product between two points on a plane, expressed in polar coordinates, is the product of their radii times the sine of the difference in their phases. Therefore, if:
\begin{equation}
    \vec{y}(x) = \sqrt{|A|}[\cos(2 \pi \text{sign}(A) \omega x), \sin(2 \pi \text{sign}(A) \omega x)]
\end{equation}
then:
\begin{equation}
\begin{aligned}
    \text{disc}(\vec{y}(x),\vec{y}(y)) & = \sqrt{|A|}^2 \sin(2 \pi \text{sign}(A) \omega (x - y)) \\ & = \text{sign}(A) |A| \sin(2 \pi \omega (x - y)) = A \sin(2 \pi \omega (x - y)) = f(x,y).
    \end{aligned}
\end{equation}

\textbf{Lemma 1: [Trigonometric Performance Functions of One Trait]} \textit{If $f(x,y) = A \sin(2 \pi \omega (x-y))$ for $x,y$ both in a one-dimensional trait space, then $f$ is disc game embeddable using the embedding:}
$$\vec{y}_k(x) = \sqrt{|A|}[\cos(2 \pi \text{sign}(A) \omega x), \sin(2 \pi \text{sign}(A) \omega x)]$$.

Notice that this construction maps the intervals of length $1/\omega$ in $\Omega$ to the circle of radius $\sqrt{|A|}$ centered at the origin. It follows that, if a performance function is embeddable onto a circle centered at the origin then there exists a mapping from trait space to the real line where performance is of the form \ref{eqn: sin performance}. 

This result extends easily to linear combinations of sinusoidal functions with varying frequencies. Consider a performance function of the form:
\begin{equation}
    f(x,y) = \sum_{k=1}^n A_k \sin(2 \pi \omega_k (x - y))
\end{equation}

Then, $f$ can be recovered using a sum of $n$ disc game embeddings, where the $k^{th}$ embedding has the form:
\begin{equation}
    \vec{y}_k(x) = \sqrt{|A_k|}[\cos(2 \pi \text{sign}(A_k) \omega_k x), \sin(2 \pi \text{sign}(A_k) \omega_k x)]
\end{equation}

%% show how to embed a 1D translationally invariant function with a sine series

Note that, all the performance functions of this kind are translation invariant since they are functions of the difference $x - y$, which does not change if after shifting $x$ and $y$ by some amount $s$. :

\textbf{Theorem 1: [Translation Invariant One Trait Performance Functions]} \textit{Suppose that $\Omega$ is a one-dimensional trait space, and $f(x,y)$ is translation invariant. Then there exists a function $h$ such that $f(x,y) = h(x - y)$. Suppose in addition that $h(x)$ is periodic with period $P$, or $\Omega$ is contained inside an interval with length $P/2$. Then $f$ is disc game embeddable using a countably infinite sequence of disc games, which correspond to the sine series expansion of $h$ and converge under the same conditions as the sine series. Moreover, each disc game represents a term in the sine series, and maps $\Omega$ to a subset of a circle centered at the origin with radius fixed by the corresponding coefficient in the sine series.}

\textbf{Proof:} If $f(x,y)$ is translation invariant then $f(x,y) = h(x - y)$ for some function $h$. Since $f(x,y) = -f(y,x)$, $h$ must be an odd function. If $\Omega$ is contained inside an interval of length $P/2$, then $h$ can be extended to an odd, continuous, $2 P$ periodic function, or an odd $P$ periodic function. If not, then, by assumption, $h$ is periodic with period $P$.

All integrable $P$ periodic functions on the real line can be approximated with a Fourier series. If the function is real valued and odd, then the Fourier series is a sine series of the form:
\begin{equation}
    h(x) \simeq \sum_{k=1}^{\infty} A_k \sin(2 \pi \omega_k x), \quad \omega_k = \frac{k}{P}.
\end{equation}

Each term in the sine series can be reproduced by a disc game embedding using the method for embedding sinusoidal functions introduced before. Specifically, let:
\begin{equation}
    \vec{y}_k(x) = \sqrt{|A_k|}[\cos(2 \pi \text{sign}(A_k) \omega_k x), \sin(2 \pi \text{sign}(A_k) \omega_k x)]
\end{equation}
where $A_k$ is the $k^{th}$ amplitude in the sine series of $h$:
\begin{equation}
    A_k = \frac{4}{P} \int_{0}^{P/2} h(x) \sin(2 \pi \omega_k x) dx.
\end{equation}

Then, a partial expansion in terms of $r$ disc games equals the $r$ term sine series expansion of $h(x-y)$:
\begin{equation}
    \sum_{k=1}^r \text{disc}(\vec{y}_k(x),\vec{y}_k(y)) = \sum_{k=1}^n A_k \sin(2 \pi \omega_k (x - y)).
\end{equation}

Thus, convergence of the sequence of disc game embeddings follows convergence of the sine series expansion. $\square$

It remains to show that, after sampling a finite set of agents, the result of PTA recovers the sine series representation. Sine series are low rank optimal in this case since, if the agents are ordered by increasing $w(x)$, the evaluation matrix is of the form: 
\begin{equation}
    F = \left[\begin{array}{ccccc} 0 & 1 & 1 & \hdots & 1 \\
    -1 & 0 & 1 & \hdots & 1 \\
    -1 & -1 & 0 & \hdots & 1 \\
    \vdots & \vdots & \vdots & \ddots & \vdots \\
    -1 & -1 & -1 & \hdots & 0
    \end{array} \right]
\end{equation}

This matrix is real, skew-symmetric, Toeplitz, and is diagonalized by the discrete Fourier transform, so PTA produces a sine series. The sequence of disc games act as the sine series expansion of a step function, with each higher order disc game corresponding to a higher order correction of an approximation to the step function. The first disc game is transitive and captures 90\% of the structure of the evaluation matrix. Subsequent disc games correct the first disc game in order to produce a step function. While it only takes two disc games to recover 95\% of the structure of $F$, so the 95\% complexity of $[1,2,4]$ blotto is 2, the subsequent eigenvalues decay slowly, so stricter accuracy requirements lead to large complexities. The slow decay of the eigenvalues is a natural consequence of the slow convergence of sine series to a step function. Here it is clear that the complexity predicted by PTA overstates the complexity of the underlying game, since subsequent disc games are best interpreted as corrections that gradually finesse the first disc game, not distinct trade-offs.

More general proof and exploration is saved for future work. 

\section{Disc Games}\label{appendix:disc_games}

\subsection{Geometry}

Principal trade-off analysis is a useful visualization technique since disc games encode performance relations via embedding geometry. Reading a disc game requires familiarity with this geometry. Namely, familiarity with the various interpretations of a cross product. Here we review some relevant relations.

Cross products are closely related to area in the embedding. Given a pair of competitors with embedding coordinates $\vec{y}_k(i)$ and $\vec{y}_k(j)$, the performance of $i$ against $j$ in disc game $k$, $\text{disc}(\vec{y}_k(i),\vec{y}_k(j))$, equals the signed area of the triangle with vertices $\vec{y}_k(i)$, $\vec{y}_k(j$, and the origin. Further, the degree of cyclicity on a loop $\mathcal{C}$ of competitors can be computed by evaluating a path sum of the advantages around the loop, i.e.~$\text{curl}(\mathcal{C})$ \citep{HHD}. The curl on loop $\mathcal{C}$ equals the signed area of the loop traced out in each embedding \citep{strang2022quad}, summed over the embeddings. It follows that curl inherits the invariances of areas. In particular, curl is translation invariant. 

The cyclic component $F_c$ on a given edge $i,j$ equals the average curl over all possible triangles formed by drawing a random $k$. Since curl is translation invariant, the cyclic component of competition is translation invariant. In contrast, the transitive component of competition is not translation invariant, and translation in a disc game induces a transitive component of competition \citep{balduzzi2018re}. 
By subtracting $F_t$ from $F$ to recover $F_c$ we center all the rows and columns, so the scatter cloud of embedded competitors will be centered at the origin. In contrast, if we embed $F$ directly, then transitivity arises from translation of each scatter cloud away from the origin. If the origin is not included in the convex hull of the embedded agents, then competition is transitive.

In contrast, scaling the embedding coordinates does change the predicted performance relations. Area is proportional to length squared, so scaling the embedding coordinates by $\sqrt{s}$ scales the associated cyclic component of competition by $s$. The scaling from $\hat{Y}$ to $Y$, was adopted to ensure that unit area in embedding generates unit curl. It follows that the area encompassed by a set of points in a disc game embedding directly represents the amount of cyclic competition among those agents, and hence the importance of that embedding.

\begin{figure}[t]
    \begin{center}
        \centerline{\includegraphics[scale=0.3]{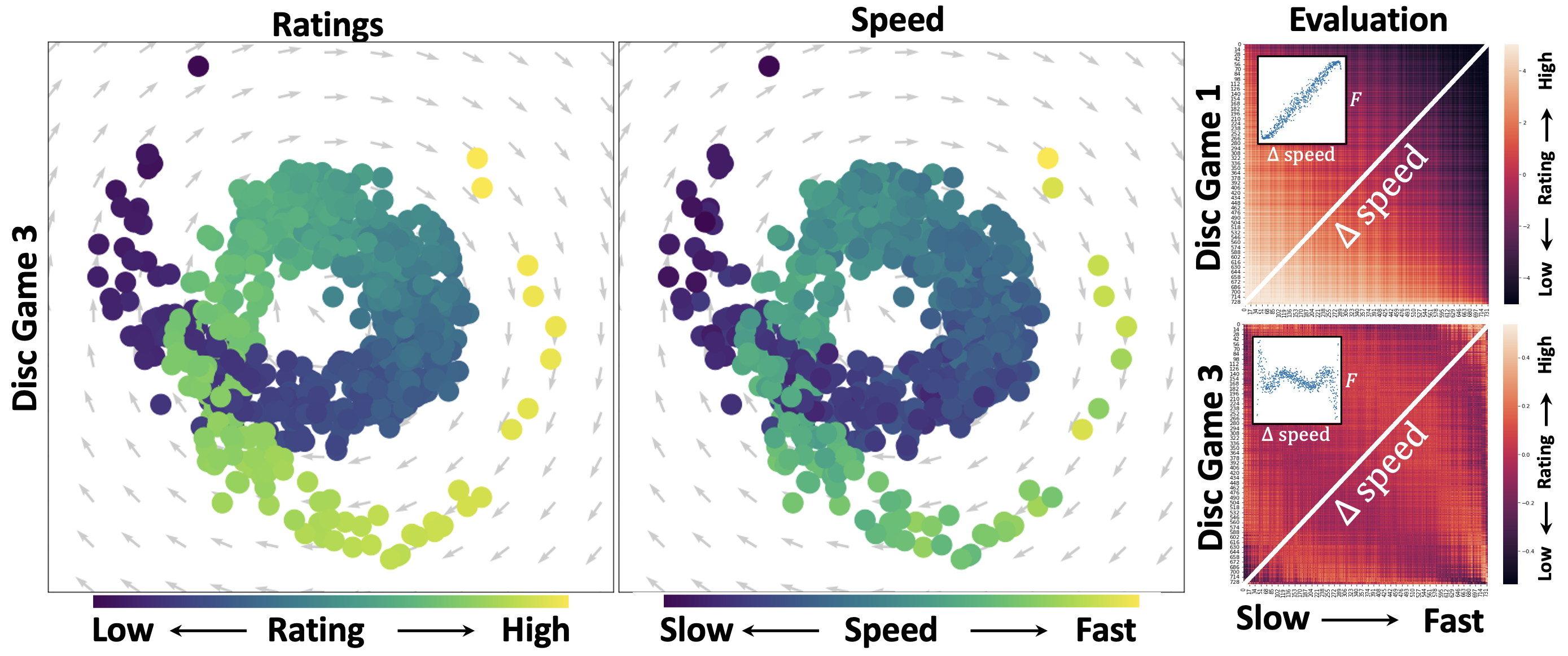}}
    \end{center}
    \caption{\textbf{Left:} Disc game 3 colored by rating. \textbf{Middle:} Disc game 3 colored by speed. \textbf{Right:} Evaluation matrices generated by disc games 1 (top) and 3 (bottom), with agents ordered by increasing speed. Both evaluation matrices are close to Toeplitz, so produce evaluation that depend primarily on the difference in speed between agents. The function that returns performance given a speed difference is approximated by sampling the evalutation matrix along the cross-diagonal marked in white. The subpanels containing scatter plots show the sampled evaluations.}
    \label{pokemon_discgame_3}
\end{figure}

\section{Pokemon}\label{appendix:pokemon}

Here we describe in further detail the construction of the Pokemon data. 
In the full game a player (or trainer) captures Pokemon to compete against the Pokemon of other players, usually with teams of 6 Pokemon chosen at the player's discretion. Players also choose the order in which their Pokemon compete since the actual combat is done pairwise. This pairwise interaction is what allowed us to ignore the team aspect and still learn important aspects of the game. Each of the pokemon had a set of attributes. The attributes are shown below. 
\newline
\newline
\newline

\begin{enumerate}
    \item Type 1: Main Type - Fire, Water, Grass, ect...
    \item Type 2: Secondary Type - Not all pokemon have two types but we did not find this to contribute to any performance tradeoffs in a significant way
    \item HP: Hit points - Indicated how much damage a pokemon can endure before losing the match. 
    \item Attack: Base modifier for normal attacks
    \item Defense: The base damage resistance against normal attacks
    \item Special attack
    \item Special Defense
    \item Speed: This stat largely determines which pokemon get to attack first. As combat is turn based, this constitues a large advantage which we saw in disc game 1. 
\end{enumerate}

There were 50,000 pairwise interactions among the 735 pokemon that were used. The data for each interaction consisted of the name of the first and second pokemon as well as the winner of the match.  In an individual matchup, each Pokemon has a certain level of HP or health. The two Pokemon take turns attacking one another until one of the them loses all of
their HP and is declared the loser. The first to attack is determined by some set of attributes that is not explicitly given by the data set, but speed is known to be a large contributing factor. Since we did not have the full interaction graphWe filled in any missing data using logistic
regression, producing a win probability matrix. We obtained the evaluation matrix $F$ via the logistic link function commonly used in Elo rating. The evaluation for competitor $i$ vs competitor $j$ is then given by $f_{i,j} = \log(\frac{p_{ij}}{1 - p_{ij}})$ where $p_{ij}$ is the probability that Pokemon $i$ beats Pokemon $j$. We applied the Schur decomposition directly to $F$ to show that disc game embedding can successfully isolate a dominant transitive component (speed).

In Figure~\ref{pokemon_discgame_3} we show the third disc game left out of the main analysis. It shows a double loop structure with a full inner circle and a half outer circle. Like disc game 1, disc game 3 is, approximately, a curve parameterized by speed. As in the Fourier examples discussed before, the double loop represents a higher order correction to disc game 1. Disc game 1 confers a transitive, monotonically increasing advantage to faster agents. The faster an agent relative to their opponent, the larger their advantage. Disc game 3 adds nuance to this relation by discounting the advantage conferred by small differences in speed, increasing the advantage conferred by intermediate differences in speed, discounting the advantage conferred by large speed differences, and strongly rewarding maximal speed differences (see the evaluation matrix and associated subpanel in the rightmost column of Figure \ref{pokemon_discgame_3}).

Note, these corrections to disc game 1 are very small. The eigenvalue for disc game 1 is roughly 15 times larger than the eigenvalue for disc game 3, hence the relationship between speed and performance is largely determined by disc game 1. We did not discuss disc game 3 in the main text for this reason.